\documentclass[manuscript]{acmart}
\AtBeginDocument{%
  \providecommand\BibTeX{{%
    \normalfont B\kern-0.5em{\scshape i\kern-0.25em b}\kern-0.8em\TeX}}}

 \acmBooktitle{Woodstock '18: ACM Symposium on Neural Gaze Detection}
\acmPrice{15.00}
 \acmISBN{978-1-4503-XXXX-X/18/06}

\usepackage{hyperref}
\usepackage{multirow}
\usepackage{blindtext}
\usepackage{tcolorbox}
\usepackage{graphicx}
\usepackage{subfigure}
\usepackage{tikz}
\usepackage{amsmath}
\usepackage{multicol}
\usepackage{soul,xspace}
\usepackage{tabularx}
\usepackage{enumitem}
\usepackage{natbib}
\usepackage{threeparttable}

\usepackage{xcolor}
\usepackage{color, colortbl}
\usepackage{longtable,tabu}
\definecolor{Gray}{gray}{0.9}
\newif\ifdraft
\draftfalse

\newcommand{\boldification}[1]{\ifdraft\indent ** \textbf{#1} **\\\indent\else\relax\fi}

\newcommand{\eg}{{\it e.g.,}\xspace}

\newcommand{\etal}{{\it et~al.}}
\newcommand{\ie}{{\it i.e.,}\xspace}

\newcommand{\ci}{{\it (i) }}
\newcommand{\cii}{{\it (ii) }}

\newcommand{\squishlist}{
\begin{itemize}[noitemsep,nolistsep]
  \setlength{\itemsep}{-0pt}
}
\newcommand{\squishend}{
  \end{itemize}
}

\begin{document}

\title [\Large \bf If This {\it Context} Then That {\it Concern}]{\Large \bf If This {\it Context} Then That {\it Concern}: Exploring users' concerns with IFTTT applets}

\author{Mahsa Saeidi  }
\email{saeidim@oregonstate.edu }

\affiliation {
 \institution{ Oregon State University}
 \country{USA}
}

\author{McKenzie Calvert}
\email{calvertm@oregonstate.edu }
\affiliation{%
  \institution{Oregon State University}
  \country {USA}
  }

  \author{Audrey W. Au}
\email{auau@oregonstate.edu }
\affiliation{%
  \institution{Oregon State University}
  \country{USA}
}

\author{Anita Sarma}
\email{sarmaa@oregonstate.edu }
\affiliation{%
  \institution{Oregon State University}
  \country{USA}
  }
  
  \author{Rakesh B. Bobba}
  \email{rakesh.bobba@oregonstate.edu }
\affiliation{%
  \institution{Oregon State University}
  \country{USA}
}

\renewcommand{\shortauthors}{}


\begin{abstract}
End users are increasingly using trigger-action platforms like, \emph{If-This-Then-That} (IFTTT) to create applets to connect smart home devices and services.
 However, there are inherent risks in using such applets---even non-malicious ones---as sensitive information may leak through their use in certain contexts (e.g., where the device is located, who can observe the resultant action). This work aims to understand how well end users can assess this risk. We do so by exploring users' concerns with using IFTTT applets and more importantly if and how those concerns change based on different contextual factors. Through a Mechanical Turk survey of 386 participants on 49 smart-home IFTTT applets, we found that nudging the participants to think about different usage contexts led them to think deeper about the associated risks and raise their concerns. Qualitative analysis reveals that participants had a nuanced understanding of contextual factors and how these factors could lead to leakage of sensitive data and allow unauthorized access to applets and data.

\end{abstract}





\maketitle

\section{Introduction}

\boldification{IoT is being increasingly used, and TA-programming platforms make it easier to use for end-users; because of Ease of IFTTT usage  millions of regular end users use these services}

Many homes are being out-fitted with Internet connected sensors and devices to create interactive and adaptive smart living spaces with a promise of convenience, safety, security, and energy efficiency. Programming platforms and frameworks such as IFTTT~\cite{ifttt}, OpenHAB~\cite{openhub}, and Microsoft Flow~\cite{micro}  enable end users to compose different smart devices and services to make it easy for them to monitor and control their smart home environments. Frameworks like IFTTT use very simple trigger-action formats--\emph{``if trigger then action”}--to enable new and rich functionality. For example, consider a simple applet that allows end users to control their smart camera via voice assistants such as Alexa; users can then voice activate their (hidden) nanny camera before going to work. ``Regular end users'' can now buy inexpensive off-the-shelf devices from different vendors and then connect them using frameworks like IFTTT. Because of this ease of creating ``applets'' that can connect devices from different vendors, and the proliferation of connected devices, \textit{millions use these services}; some simply reusing already created applets, others creating their own. For IFTTT alone, there are \textit{18 million registered users}, running over 1 billion IFTTT applets (originally called recipes) each month in 2020~\cite{IFTTT-stats-2020}.

\boldification{despite of ease of use, there are risks associated with these services}

While these platforms are easy to use and allow new functionalities, composing different services or connecting devices to services can lead to unexpected or undesirable behavior~\cite{celik2018soteria}, especially if not deployed thoughtfully. For instance, in our example, a babysitter may simply turn off the nanny camera with a voice command to Alexa (since Alexa doesn't use any kind of filtering through voice recognition) without the user's knowledge (unauthorized modification). In this case, the voice assistant was in an accessible location and had no authentication capability undermining safety for the baby.

As another example, consider an applet that connects a smart camera to an online photo-storage service by taking and logging a picture when the camera detects motion. While this is a useful applet that can be used for home monitoring while one is away, deploying this applet can have severe privacy consequences depending on where the camera is located (\eg living room or bedroom), when it is active (\eg during working hours vs. all day), and who else has access to the online photo storage (\eg private folder vs. shared folder). 

\boldification{the popularity of IFTTT and the associated risks bring us to the RQ1, Do people understand the risks of deploying such applets}

Given the popularity of IFTTT applets and the potential risks associated with deploying them, we wanted to see if and to what extent people think about these risks when considering an applet. This brings us to:
\textit{RQ1: How concerned are users about using IFTTT applets?} Especially, when given a simple description of the applet that explicitly identifies the applet \emph{trigger} and \emph{action}.

\boldification{Further Risks are Context dependent }
Further, as prior examples show, the risks with applets can be context dependent. For instance, in the home monitoring example above, logging smart camera pictures into a shared folder poses privacy risks, but the risk can be significantly different depending on the camera's location. If the camera is on the front porch, a more-or-less public location it is less of a concern as opposed to it being in the living room or other more private areas in the home. Time of the day matters too, pictures taken during the daytime may be less sensitive than those of individuals in their night attire (in the privacy of their home). This highlights the fact that the context of use can impact the risk of using such applets, and thus, a more nuanced view of context may help to identify these risks better. Therefore, we wanted to see if and how contextual factors influence users' concerns bringing us to: \textit{RQ2: To what extent do contextual factors impact end-users' concerns?} 

Given that millions of end users are downloading and using IFTTT applets, without a deeper understanding of end users' current concerns, especially to what extent these concerns are impacted by the context of use, approaches to help users understand or mitigate the risks associated with home automation will have limited success.   
 
For instance, without considering the context of use, current efforts that automatically identify potential risks with applets by tracking the flow of information from (public or private) sources to sinks (\eg ~\cite{surbatovich2017some, bastys2018if}) may be inadequate. 

While previous research in traditional mobile applications (\eg ~\cite{sadeh2009understanding,lin2012expectation}) or in sharing of online content (\eg ~\cite{habib2019impact}) has investigated the impact of context, their findings are not directly applicable to home automation. This is because the interaction between physical environment and (multiple) sensors and devices in a smart-home environment leads to a unique, more nuanced, and richer context than previously considered.

\boldification{therefore to close this gap in our understanding of users' concerns we did online study}

To close this gap in our understanding of users' concerns with IFTTT applets we conducted an online Mechanical Turk based survey with 386 participants, who in total answered questions about 49 popular IFTTT applets. The survey first asked participants whether they would be concerned about using an applet by presenting them with a simple description (RQ1). It then specifically prompted participants to think about the contextual factors relevant to that applet (RQ2) through a subsequent set of questions.

Our findings indicate that nudging users to think about different usage contexts can help them better appreciate risks with applets and each contextual factor significantly impacts users' concerns with using these applets. Further, in a majority of cases, participants were concerned because of security and privacy risks associated with the context of use of these applets.

\section{Related Work} 

Our work aims to understand how well end users are able to assess potential risks with using \emph{trigger-action} applets in smart home environment. Further, we investigate how contextual factors might influence their assessment. Hence we focus on related work on the impact of context on users' concerns in mobile and Internet-of-Things (IoT) environments. We also survey work that looks at the risk from the perspective of unintended information flows and unauthorized accesses in IoT frameworks for smart homes.

\subsection{Impact of Context on Users' Concerns}

\noindent \textit{Smart Home and IoT:}
Much work has been done to understand users' privacy preferences broadly in IoT applications and frameworks. Importantly, \ci Emami-Naeini \etal~\cite{naeini2017privacy} investigated users' comfort level with data collection scenarios, and \cii Lee \etal~\cite{lee2016understanding,lee2017privacy} explored the factors that affect users' privacy preference in IoT. Other studies investigated what causes users to accept home sensing systems~\cite{choe2012investigating} and factors that influence users' preferences about giving others access permissions to use the IoT apps~\cite{he2018rethinking}. 

In this work, we focus on users' concerns with real world trigger-action applets that connect two or more IoT devices/services rather than on individual IoT devices~\cite{lee2017privacy,lee2016understanding,he2018rethinking} or IoT data collection scenarios~\cite{naeini2017privacy}. We use different contextual factors that are relevant to the usage of the applets in the home environment rather than those related to data collection~\cite{naeini2017privacy}. We also considered both \emph{who can trigger} the applets and \emph{who can observe} the applet actions as contextual factors which are different from those in~\cite{he2018rethinking}. Further we differ from all previous works as we investigate how users' perceived concerns with different applets evolve after being given just the applet description, and after being presented with different contextual factors.

\noindent \textit{Mobile Applications:}
Much work on evaluating factors influencing users' concerns with mobile apps also exists. For example, Sadeh \etal~\cite{sadeh2009understanding} 
found that increasing users' awareness would help them make better choices regarding sharing their location information. Lin \etal~\cite{lin2012expectation} 
found that giving information about why a resource is being used by a mobile app can impact users' privacy concerns related to mobile apps. Tsai \etal~\cite{tsai2009s} 
found that users became more comfortable with sharing their location data with friends and strangers after seeing feedback about who has viewed their location data, and when the location data was shared. In contrast, we investigate users' concerns with using applets that connect multiple home IoT devices or online services.

\subsection {Security and Privacy Concerns in IoT }

\noindent \textit{IoT programming frameworks:} The coarse-grained permission models that are being used by current programming frameworks (\eg Samsung SmartThings~\cite{SAMSUNG}, Apple's HomeKit~\cite{apple}, OpenHAB) are ineffective in controlling sensitive information flows and unauthorized access of sensitive data. Fernandes \etal~\cite{fernandes2016security} evaluated SmartThings and discussed design flaws that lead to over-privileged apps. Several solutions to mitigate sensitive information flows and unauthorized accesses within IoT apps have also emerged. For example, ContextIoT~\cite{jia2017contexlot}, is a context-based permission system that identifies the usage context of sensitive actions using control and data flow information. Similarly, SAINT~\cite{celik2018sensitive} uses static analysis of application code to identify sensitive information flows between taint sources and taint sinks. SmartAuth~\cite{tian2017smartauth} system uses device information and app descriptions to identify over-privileged apps. Those approaches aim to fix sensitive information flows and improve access control models within a single app. 
However, explicit and implicit interactions between multiple IoT apps could also lead cause confidentiality and integrity violations. 

Approaches to address cross application violations have also been proposed. For example, FlowFence~\cite{fernandes2016flowfence} approach enforces information flow controls by restricting usage of sensitive data inside sandboxes. 
ProvThings~\cite{wang2018fear} framework uses provenance data to identify malicious information flows across security sensitive IoT apps and device APIs. SOTERIA~\cite{celik2018soteria} uses the state model of individual or set of apps to check safety, security, and functional properties and find property violations. IoTGuard~\cite{celik2019iotguard} proposes a dynamic policy-based enforcement system to protect users against integrity and confidentiality violations using predefined safety and security policies on individual or set of interacting apps.

In the preceding approaches, low level system contextual information including source code and/or security labels for data and entities defined by security experts were used to identify information flow-based and other security violations. However, such approaches cannot capture security and privacy violations that are dependent on high level semantic contextual factors such as usage context, and user privacy perceptions and preferences. Our findings in this work show that high level (as opposed to low level system context) contextual factors play an important role in users' perceived risks and hence in defining what constitutes a risk.

\noindent \textit{Trigger-action programming frameworks:} Emerging trigger-action programming frameworks (\eg IFTTT, Microsoft Flow, Zapier~\cite{zapier}) that help end users to connect IoT devices with online services, also suffer from weak access control and  other security and privacy issues similar to IoT programming frameworks. Lack of fine-grained access controls can lead to privacy and integrity violations~\cite{zhang2017dolphinattack}. Surbatovich \etal
~\cite{surbatovich2017some} proposed a security lattice model that uses labeled triggers and actions to identify sensitive information flows in IFTTT. A similar approach was used by Bastys \etal~\cite{bastys2018if} to label triggers, actions, and to automatically prevent identified violations (\eg integrity, confidentiality, availability) by breaking the information flows from private sources to public sinks. The labels that are used in these previous two approaches are coarse grained and are based on \emph{who can use} and \emph{who can observe} the data and do not consider other usage contexts. For instance, those approaches do not differentiate between the different people that might be able to use the applets or observe them. 

In another recent work~\cite{cobb2020risky}, Cobb\etal~ used similar coarse-grained labels to manually investigate violations and asked users to react to specific confidentiality and integrity violations discovered in a set of applets. They recognized that granularity of secrecy and integrity labels can be improved through a better understanding of contextual factors. Our work not only empirically identifies that contextual factors matter to participants, but also analyzes the importance of these factors and their interplay. We also uncover other areas of concerns (e.g. safety) depending on the context that might be considered when creating tools to help end users better understand impacts of using IoT devices.

The frameworks and systems discussed here can benefit from our work as they can be improved to create more fine-grained information flow policies based on our findings.

\section{Methodology} 
\label{sec:method}

We conducted an online survey, approved by our Institutional Review Board (IRB), to explore users' concerns with IFTTT applets.  
In the survey, we first asked participants about their concerns with using an applet after reading a simple trigger-action description (without any potential contextual information in the original description). Next, the survey presented participants with a set of contextual factors and asked them about their concerns with using the applet for each context. We also collected demographic data including age, gender, education, and IoT/IFTTT related background.

\begin{table*}[ht]
  \begin{center}
    \caption{Demographic of participants}
    \label{tab:demographics}
 \small
 \setlength\tabcolsep{3pt} 
    \begin{tabular}{p{1.5cm} p{.9cm}| p{1cm} p{1cm} |p{2.7cm} p{1cm}| p{.5cm} p{.9cm}| p{.5cm} p{.9cm}}
    \hline
    \multicolumn{2}{c}{\textbf{Gender}} & \multicolumn{2}{c}{\textbf{Age}} & \multicolumn{2}{c}{\textbf{Education}} & \multicolumn{2}{c}{\textbf{IFTTT applets}}  & \multicolumn{2}{c}{\textbf{IoT devices}} 
    \\
      \hline
      Women &   $39\%$ & $18-24$ & $10.4\%$ & Less than high school & $0.25\%$ & 0 &$22\%$&0&$10.1$\% 
      \\
      
      Men &$60\%$ & $25-34$ & $60.1\%$ & High school   & $11.9\%$ & $1-2$& $61\%$&$1-2$& $65\%$
      \\
      Non-binary &$<1\%$ &$35-44$ &  $21.5\%$ & Some College   &$13\%$ & 3-4&$11.14\%$ & 3-4&$18.4\%$ 
      \\
      Trans & $<1\%$ &$45-54$ & $4.15\%$&2 year degree&$10.4\%$& 5+&$6.74\%$&5+&$6.48\%$
      \\
      & &$55-65$ &$3.37\%$& 4 year degree&$51.8\%$&&&&
      \\
      & &$65+$ & $0.51\%$&Professional degree&$11.9\%$ &&&&
      \\
      & & &&Doctorate&$0.77\%$ &&&&
       \\
      \hline
    \end{tabular}
   
  \end{center}
      \vspace{-1.5\baselineskip}
\end{table*}

\subsection{Recruitment and Participants}
We recruited $386$ participants through Amazon Mechanical Turk (MTurk). 
Participants had to be at least 18 years old, live in the United States, have an approval rate of $95\%$ or greater, and have at least 100 HITs approved. We compensated participants $\$3.50$, as our pilot study with 4 participants showed the survey took between 20 to 30 minutes. The MTurk participants took an average of $24$ minutes. 
Of the $386$ participants, $60\%$ identified as men, $39\%$ identified as women, and less than $1\%$ from other categories. Majority of participants owned one to two IoT devices ($65\%$) and used one to two IFTTT applets ($61\%$). Majority of participants had some college education or higher ($88\%$). Table~\ref{tab:demographics} captures demographics of the participants.

\subsection{Study Design}\label{subsec:studydesign}
 
\noindent \textit{\textbf{ Applet selection. }}\label{subsec:IFTTTapplet}
We selected the 50 most frequently used applets from each smart home related category from the IFTTT dataset published by Ur \etal~\cite{ur2014practical}. The categories we used were ``appliances", ``lighting", ``environment control \& monitoring", ``security \& monitoring systems", ``location", ``smart hubs \& systems", and ``voice assistants". This gave us a total of $350$ applets.
We reviewed all applets' descriptions and removed the following types of applets: (1) duplicates, (2) those that did not have clear descriptions, (3) those that involved services that are now uncommon (e.g., Ubi), (4) those that were not related to smart homes (e.g., News applets), and (5) those that were IFTTT specific services, such as sending notification to IFTTT e-mail. After this filtering stage we had $90$ applets. Next, two researchers discussed each applet to identify if there were potential security and privacy risks for each applet. As our goal was to understand end-users' concerns in using potentially risky applets we filtered out those applets for which the researchers could not come up with potential risks. This resulted in a final dataset of $49$ applets\footnote{The list of applets' descriptions are provided in the Appendix.}.
 
Finally, we converted all applet descriptions to a standardized format that explicitly stated the trigger and actions. We did this because in some cases the descriptions in the dataset were confusing or had very little description of the trigger or action. For instance, for the applet that blinks the Hue light when the user receives a new SMS  (Applet\#20), the IFTTT description was:\textit{``Never miss an important text on your Android phone with this Applet."} The standardized format helped us direct participants to think about the trigger-action rule.

\noindent \textit{ \textbf{Context description.}} We selected those contextual factors that are related to home environments. We included ``location" as a factor as it has been found to be relevant in prior work (\eg~\cite{naeini2017privacy,lee2016understanding,lee2017privacy,he2018rethinking}). We considered both the location of the \emph{triggering} IoT device and the location of the resultant \emph{action}. Since IFTTT applets that connect IoT devices to online services are an important class, we explicitly considered their action location, \ie action in online services as a separate contextual factor. Further, we analyzed the situations in which integrity and confidentiality violations might occur~\cite{surbatovich2017some}. This led us to consider \emph{who is involved} in triggering and \emph{who is observing} applets as contextual factors. Finally, we considered the ``time of the day" as it can lead to different security and privacy implications. We now discuss briefly the list of values for each contextual factor in our study:

\squishlist
\item {\bf Trigger Location} considers the area in the home where the IoT device is located (trigger events occur). We considered public areas as places that are accessible to outsiders (e.g., front entrance areas), semi-private areas as those that are accessible to visitors and homeowners (e.g., living room and kitchen), and private areas as those where access to them is largely limited to homeowners (e.g., bedroom and bathroom).

\item {\bf Action Location} considers the area in the home where the IoT device or service that acts as an action service is located. It included public areas such as,  front entrance areas, semi-private areas such as, living room and kitchen, and private areas such as, bedroom and bathroom.
\item \textbf{Action in Online Services} includes situations where the applet's action is reported/logged in online services (e.g., a Facebook post when the user arrives home). It captures concerns where the information through online services is accessible to (or shared with) others. For example, if the Facebook post is on a public page there might be privacy concerns.

\item \textbf{Who Can Use} considers who is able to use or trigger the applet. It primarily considers the different groups of people based on their relationship to the homeowner, and included: spouse, kids, visitors, and outsiders (anyone outside of the house).

\item \textbf{Who is Around (can observe)} considers who is nearby and can observe the applet action. This factor included people based on their relationship to the homeowner: spouse, kids, visitors, and outsiders.

\item \textbf{Time} of the day that an end user uses the applet included factor values: morning, afternoon, and night.
\squishend

\noindent\textit{\textbf{Survey design. }} We designed seven surveys (7 MTurk HITS), with two sections each. The first section, common across all the surveys, covered demographic questions. We collected demographic information of participants including age, gender, education, number of IoT devices they own (or had owned), and the number of IFTTT applets they use (or had used). The second section covered applet-specific questions with each survey (HIT) covering seven unique randomly-chosen applets out of the $49$ (7 surveys x 7 applets). The demographic questions and the questions for one applet are shown in the Appendix. Each survey was released in MTurk sequentially and each participant could have responded to only one survey (HIT).

\boldification{RQ1}
\noindent \textit{Users' Concerns with applets from applet descriptions (RQ1).} 
For each applet, we asked participants to rate their concerns through a direct question: ``Would you be concerned about using this applet?". 
The answer choices were on a Likert scale ranging from \emph{Not at all concerned} (1) to \emph{Extremely concerned} (5). To avoid priming participants to specific risks, we used a general term ``concern''.

\boldification{RQ2}
\noindent \textit{Impact of contextual factors on users' concerns (RQ2).}
Then we presented participants with the context questions to find which contextual factors might influence users' concerns with using the applets. Specifically, we asked participants to rate their concerns with using each applet under the six previously discussed contextual factors.  
Similar to the previous question, participants had to choose from among the five options on a Likert scale ranging from \emph{Not at all concerned} (1) to \emph{Extremely concerned} (5). If participants' responses were at level 3 (\emph{Somewhat concerned}) or above, then we asked participants to provide an explanation of their choice 
via open-ended questions. We used participants' responses to these open-ended questions to understand if and how their concerns changed after considering contextual factors, and how their concerns differed across different contextual factors for each applet. In the rest of the paper, we refer to these responses as ``open-ended responses."

\subsection{Study Limitations}

Like any empirical study, our work has limitations. Most of our study participants were educated (4 year degree $51.8\%$) and young ($25-34, 60.1\%$), a common demographic of MTurk workers.
While a past study~\cite{kang2014privacy} had found that MTurk workers were not representative of US population, a more recent (2019) study~\cite{redmiles2019well} has shown that MTurk workers are more representative of the U.S. population to study privacy attitudes when compared to Telephone-based or Web samples. 
Further, our dataset included 49 applets across 7 categories relevant to smart homes. This is but a small set in the universe of IFTTT applets.

\subsection{Data analysis}

We report our findings based on both quantitative and qualitative analysis. 
For quantitative analysis, we converted all categorical Likert scale variables to numerical variables. We used descriptive
statistics (e.g., frequencies, means) and statistical analysis for comparisons. For statistical comparisons, we used a generalized linear mixed model (GLMM) with the random intercepts per participant and per applet to account for within-user data dependencies ~\cite{lme4}. 
The goal was to determine if a contextual factor (independent variable) was a significant factor in participants' concerns.
To investigate this, for each contextual factor, we created two regression models: (i) the \textit{null model} without any contextual factor as a fixed effect and (ii) the model with  a contextual factor as a fixed effect (called \textit{[contextual factor]-model}). Then, we used the Likelihood Ratio Test (LRT) to test the difference between the	likelihood	of	these two models. If the result was significant, then we can conclude that the contextual factor was significant in participants' concerns. If found significant, we next investigate how the different values of the contextual factor (e.g., public, semi-private, private values in Trigger Location) played a role. We ran pair-wise comparisons across these values through post-hoc tests using \emph{glht} function, which uses Tukey to account for multiple comparisons. We used a significance level of $\alpha=0.05$. All analysis were implemented in statistical software package R~\cite{Rcore} with lme4~\cite{lme4} to perform GLMM.

For qualitative analysis of the reasons behind participants' concerns, we performed inductive coding of the open-ended responses. We first evaluated the 6038 responses across all contextual factors to remove those responses that were meaningless for analysis---those that were reiteration of the Likert option (e.g., ``not extremely concerned"), not valid reasoning (anything less than 4 words), or applet definitions copied from the Internet. This left us with 2512 valid responses.
 
For each contextual factor, the lead author performed open coding on a small subset of the responses (5\% of responses chosen in random order) to create an initial codebook. Two researchers then used the codebook to independently code the responses, and discussed and updated the codebook as needed. Once the codebook was stabilized and the researchers attained a high inter-rater reliability (Cohen's $\kappa$ \cite{fleiss2013statistical}) in 20\% of the responses (Trigger Location: Cohen's $\kappa=0.85$, Action Location: Cohen's $\kappa=0.93$, Action in Online Service: Cohen's $\kappa=0.9$, Time: Cohen's $\kappa=0.9$, Who Can Use: Cohen's $\kappa=0.86$, Who is Around: Cohen's $\kappa=1$), the second researcher independently coded the rest of the responses.

\section{Users' concerns with applets (RQ1)} 
\begin{figure}
  \begin{minipage}[t]{.4\linewidth} 
 
  \includegraphics[width=\linewidth, height=5cm]{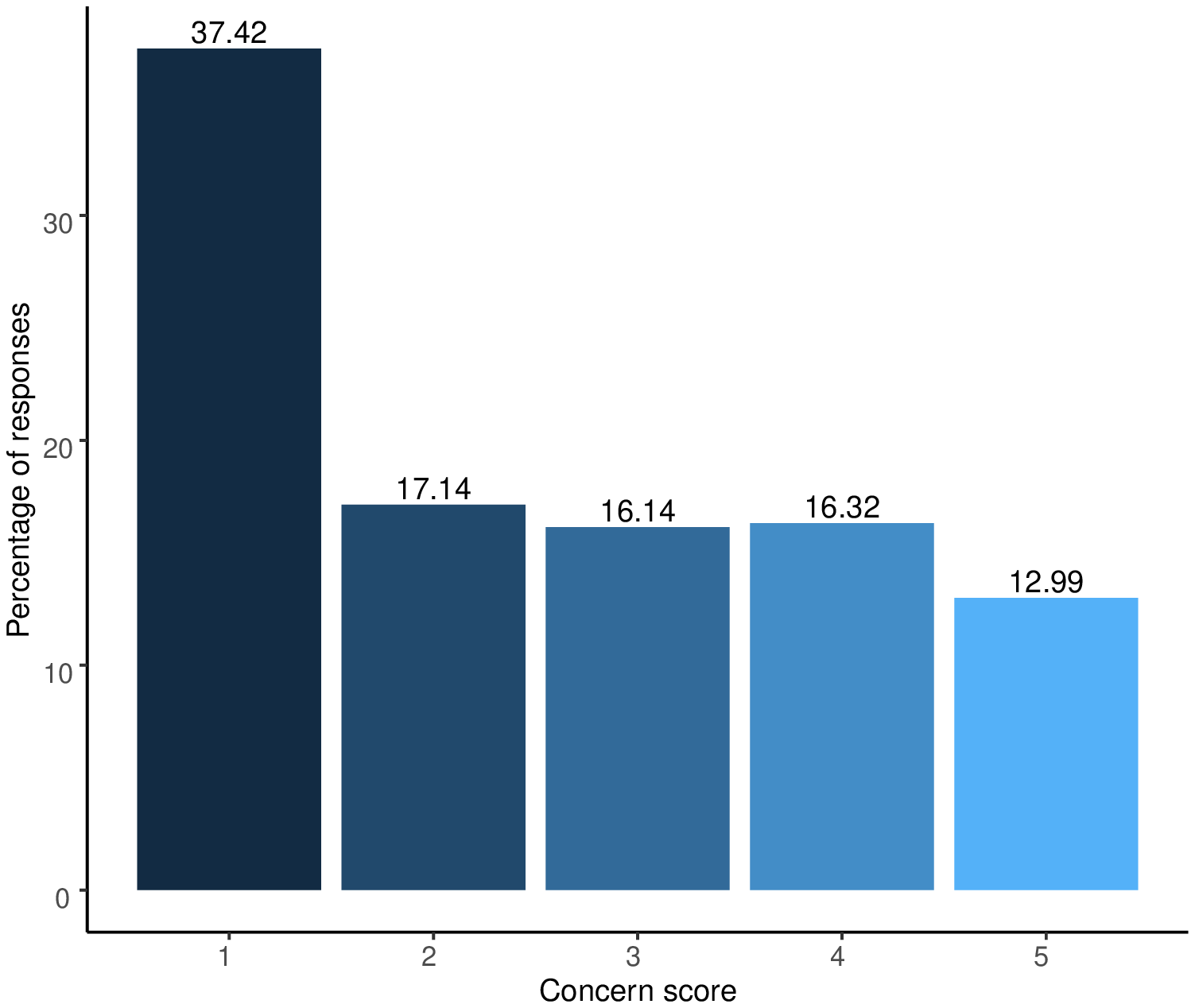}
  \caption{Distribution of participants' concern score, after reading applet descriptions across all applets (1:Not at all concerned, 2:Slightly concerned, 3:Somewhat concerned, 4:Moderately concerned, 5:Extremely concerned) }
  \label{"Barplot"}
  \end{minipage}
    \hfill
  \begin{minipage}[t]{.55\linewidth}
    \includegraphics[width=\linewidth,height=5cm]{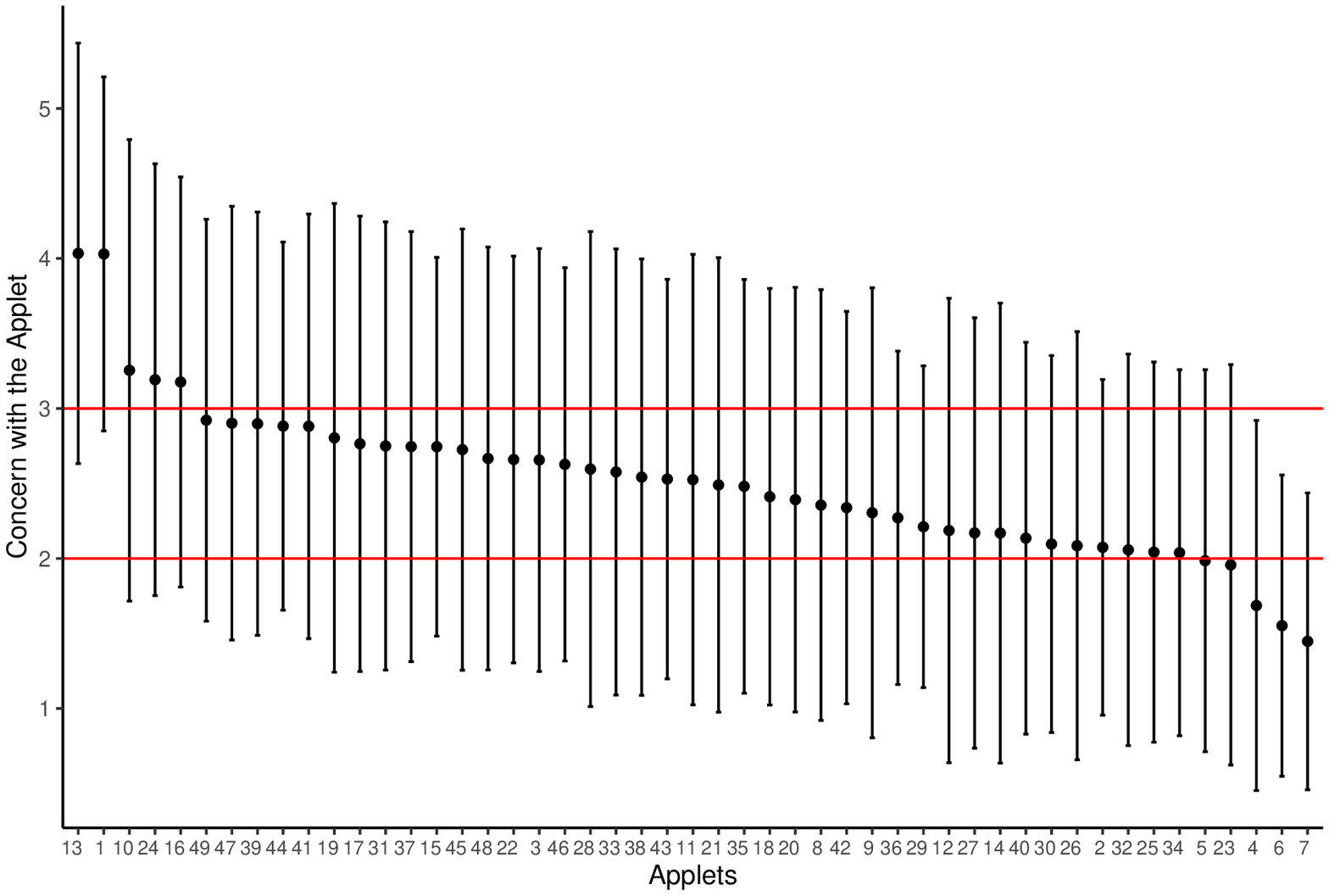}
    \caption{Average of participants' concern score after reading applet descriptions, arranged in descending order (1: Not at all concerned, 2: Slightly concerned, 3: Somewhat concerned, 4: Moderately concerned, 5: Extremely concerned)}
    \label{"mean"}
  \end{minipage}
\end{figure}

\noindent 
 
 The goal of our first research question is to create a baseline understanding of whether users have any concerns in using applets after reading their descriptions. 
This baseline allows us to gauge to what extent participants understand the consequences of using the applet when provided with a simple applet description that explicitly stated the trigger and action.

 In general, participants were not overly concerned about using the applets after reading their description. 
\emph{Not at all concerned} was the most frequently selected ($37.42\%$) and \emph{Extremely concerned} was the least selected choice ($12.99\%$), with the rest of the options more or less evenly split (around $17\%$) (See Figure~\ref{"Barplot"}).
The mean concern score was $2.5$ out of $5$ ($STDEV=1.45$), which falls between \emph{Slightly concerned} and \emph{Somewhat concerned}. 
These concerns scores are despite the fact that all applets selected for the study had a potential privacy or security risk (See Section~\ref{subsec:IFTTTapplet}).

Figure~\ref{"mean"} shows the distribution of participants' concerns for each applet in the study; arranged in a descending order of participants' concerns.
There are a few applets ($\#13$, $\#1$, $\#10$, $\#24$, $\#16$) for which participants' concern was higher than \emph{Somewhat concerned} ($3$).
It turns out that all these five applets connected a location service to an online service. For example, Applet $\#13$ connected a user's location with Facebook by posting a status update whenever the user entered a specified area. Similarly, Applet $\#1$ connected a user's location with Twitter by posting a tweet whenever the user entered a specified location. Further, all applets in our dataset that explicitly connected a user's location to an online service ($\#1,\#3,\#10,\#13,\#16, \#19, \#24, \#41$) show up in the top $20$ applets when ranked by average concern scores. In fact, $7$ out of those $8$ show up in the top $12$. We found that the means were different for these 8 applets (mean = 3.26 STDEV =1.49) as compared to the rest (mean = 2.34 STDEV =1.39) indicating that participants were concerned about their location privacy.

Although participants were concerned about leaking their location data to online services, they were less concerned when such a leak could occur indirectly. For example, participants did not seem overly concerned about using applets that connected thermostats ($\#30$, $\#46$) or light switches ($\#2$, $\#12$) to online services, which might indirectly leak participants' location or their presence at home. A case in point is Applet $\#30$, which sends an email when the thermostat is set to ``away". If someone gains access to the email logs they can infer when the participants are at home (or away). The average concern for Applet $\#30$ was between \textit{Slightly} and \textit{Somewhat concerned}. This is likely reflected in participants lack of concern for Applets $\#4$,$\#6$,$\#7$ (right end of Figure~\ref{"mean"}), which all relate to connected services/devices in a home.

We posit that this might be because the applet descriptions were insufficient in allowing participants to adequately assess the consequences of using the trigger-action applets.

\begin{figure}[t]
  \centering
  \begin{tabular}{@{}l@{}}
  \includegraphics[width=0.8\linewidth]{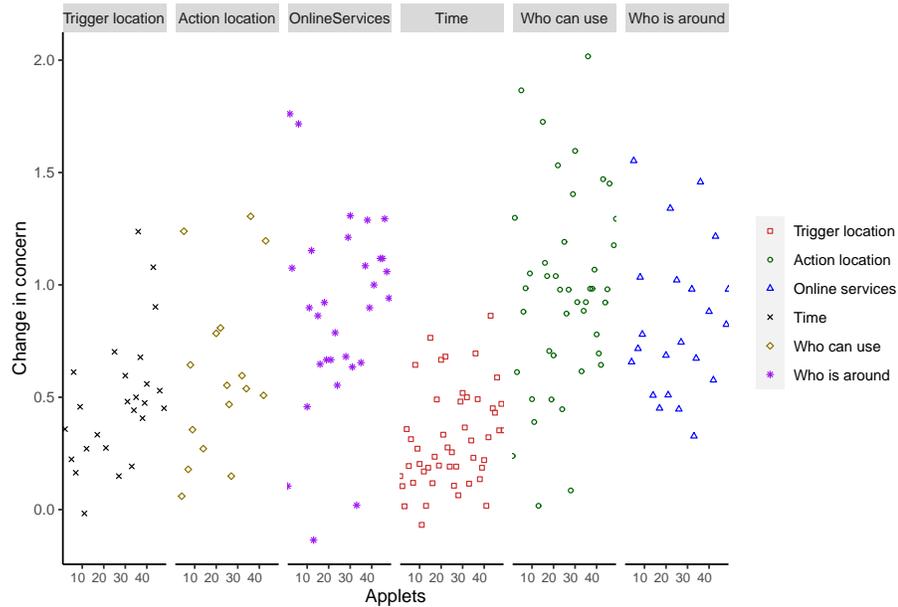}\\
  \end{tabular}
    \vspace{-.5\baselineskip}

  \caption{Average change in participants' concern when considering contextual factors. Change in level of concern score was calculated by taking the highest concern score across different values of each contextual factor (for each response)}
    \label{"Delta"}
    \vspace{-\baselineskip}
    
\end{figure}

\section{Impact of contextual factors on users' concerns (RQ2)}
 
 
 To understand the role of contextual factors, we directed participants' attention to the six contextual factors relevant to each of the 49 applets in our study. Our findings show that each and every participant---on thinking about specific context of use---\textit{increased} their concern for at least one of the contextual factors. Figure~\ref{"Delta"} shows the average change in participants' level of concerns for each contextual factor and for each applet.

As Figure~\ref{"Delta"} shows, on average, concerns increased or remained the same with few exceptions.  
Further, Figure~\ref{"Delta"} shows that the increase in concerns was higher for some applets and contextual factors than others.
To understand the details of how context impacted participants' concerns with using these applets, we conduct both qualitative and quantitative analysis.

\subsection{How significant is impact of each contextual factor}

\noindent We used regression analysis to investigate the impact of each contextual factor on participants' concerns with using the IFTTT applets. However, before running the regression models, for each contextual factor we removed the applets for which that contextual factor was not applicable.   
For instance, the contextual factor Trigger Location is not applicable for applets that activate based on the time of the day (Applet\#32 turning Phillips Hue Light on or off at specific times).

Next, we report the findings from our statistical analysis for each contextual factor categorized into three broad classes---\textit{location} of the trigger \& action, the \textit{time} of the day, and \textit{access} to the trigger \& action.

\subsubsection{Location} \label{subsubsec:loc} The location of where the applet is triggered (Trigger Location), as well as where the action is recorded---in the home or online---can have associated risks.  

\noindent\textit{\textbf{Trigger Location.}} To analyze the impact of different Trigger Locations on participants' concerns with using an applet, we categorized those applets were Trigger Location was relevant (33 out of 49 applets)
into two groups: 1) applets whose locations can only be inside the house; for example, in an applet that connects Alexa to other IoT devices, Alexa is unlikely to be located at the front door, and 2) applets that can be located both inside and outside; for example, for an applet that connects a camera to online services, the camera can be located inside or outside the house. 
Then for applets in each category, we tested the effect of Trigger Location on participants' concerns by fitting two regression models: null model and TriggerLocation-model and using LRT to assess the differences between these two models (recall from Section~\ref{sec:method}).
The Trigger Location impacted participants' concerns for applets in both categories (Significant differences for Category1: $\chi^2(df=1)=10.94, p<0.001$ and for Category2: $\chi^2(df=2)=6.13, p < 0.05$).

In the first category, participants differentiated between private (bedroom, bathroom) and semi-private (kitchen, living room) locations ($p <0.001$, See TriggerLocation-model-1 in Table ~\ref{tab:glmm}). As expected, participants were less concerned about the semi-private locations as compared to private locations (negative coefficient in Table ~\ref{tab:glmm}). In the second category, which included three types of locations (private, semi-private, public), we ran pair-wise comparisons using the \textit{glht} function to see if participants differentiated between these types of locations. Table~\ref{tab:TLglht}\footnote{The details of the pair-wise comparisons for the remaining contextual factors are in the Appendix.} shows significant differences between public and semi-private locations ($p<0.05$).

\begin{table}[t]
  \begin{threeparttable}

    \caption{Summary statistics \& regression results for contextual factor influence on participants’
concerns. We used a separate regression model for each contextual factor. First factor in each model was considered a base (intercept) by GLMM.
}
\label{tab:glmm}
 \small
 
    \begin{tabular}{
    l|cccc|l|cccc }
      \hline
 \multicolumn{5}{l}{\textbf{TriggerLocation-Model-1(inside only):3736 observations \tnote{1}{}}} & \multicolumn{5}{l}{\textbf{Time-Model: 8106 observations}}\\
\hline
     \textbf{  } & \textbf{Coeff.} & \textbf{Std.Err.} & \textbf{z-value} &\textbf{ P-value} & \textbf{ } & \textbf{Coeff.} & \textbf{Std.Err.} & \textbf{z-value} &\textbf{ P-value}  
      \\

Private locations & 
-&  
-&  
-&
-&Morning&
-&  
-&  
-& - 
\\
 Semi-private locations & $-0.071$&    0.021 & $-3.33$6  &\textbf{0.0008} &Afternoon & 0.007&  0.017&   0.452 &   0.652\\

&&&& &Night &     0.072&   0.017&   4.189& \textbf{2.81e-05}\\
\hline
\multicolumn{5}{l}{\textbf{TriggerLocation-Model-2(inside/outside):2330 observations) \tnote{2}{}}} & \multicolumn{5}{l}{\textbf{WhoCanUse-Model: 9420 observations}}\\
\hline
     \textbf{ } & \textbf{Coeff.} & \textbf{Std.Err.} & \textbf{z-value} &\textbf{ P-value}  & \textbf{ } & \textbf{Coeff.} & \textbf{Std.Err.} & \textbf{z-value} &\textbf{ P-value}
      \\

Private locations   &    
-&
-&-
&-
&  Spouse& 
-& 
-& 
-&
-\\
 Semi-private locations& $-0.03821$ &    0.03005 &  $-1.271$  &  0.204 & Kids  & 0.0491& 0.018&  2.593  & \textbf{ 0.009} 
       
      \\
 Public locations  &  0.05193 &   0.03583&  1.449  &  0.147 & Visitors &   0.188&    0.018&  10.278 &  \textbf{<2e-16}\\

      & & &  & &
      Outsider &  0.294& 0.0179& 16.428&   \textbf{<2e-16}  \\
 
 \hline
  \multicolumn{5}{l}{\textbf{ActionLocation-Model: 2400 observations}}& \multicolumn{5}{l}{\textbf{WhoIsAround-Model: 48120 observations}}\\
\hline

     \textbf{ } & \textbf{Coeff.} & \textbf{Std.Err.} & \textbf{z-value} &\textbf{ P-value} & \textbf{ } & \textbf{Coeff.} & \textbf{Std.Err.} & \textbf{z-value} &\textbf{ P-value} 
      \\

  Private locations  &  
  -&    
  -&- 
  & -
  & Spouse & 
  -&    
  -& 
  -& 
  -\\
  
Semi-private locations & $-0.12303$ &   0.02929 & $-4.201$ &\textbf{2.66e-05}& Kids  & 0.037   & 0.027   &1.367 &0.171 \\ 
Public locations   &0.02193  &  0.03450  & 0.636  &  0.525 & Visitors &0.102& 0.026  & 3.815& \textbf{0.0001} \\
&&&&&Outsiders &0.206& 0.026  & 7.883& \textbf{3.18e-1}\\
\hline

    \end{tabular}
 
      \begin{tablenotes}
     \item[1]\small{
     Applets whose triggers can only be located inside the house.}
     \item[2]\small {Applets whose triggers can be located both outside and inside.}
   \end{tablenotes}
 
      \end{threeparttable}

\end{table}

\noindent\textit{\textbf{Action Location.}}
As our goal was to compare differences in concerns across different locations that an action could be seen, we refined the subset of applets to those where the devices could be placed at different locations. This, therefore, excluded applets such as those that connect Alexa to the oven (Applet\#17) or to the printer (Applet\#7), since the oven can only be in the kitchen and the printer is unlikely to be at the front door. This resulted in a set of 9 applets that we investigate further.

The regression models and LRT results show that Action Location had a significant effect on participants' concerns ($\chi^2(df=2)=23.748,p <0.001$). According to the regression result (ActionLocation-model in Table~\ref{tab:glmm}), participants were less concerned with the semi-private locations compared to private locations. Pair-wise comparisons also confirm this distinction and also show that there was a statistical difference in participants' concerns for semi-private locations (kitchen, living room), and front door area ($p<0.05$, See Table~\ref{tab:alocglht} in the Appendix).

\noindent\textit{\textbf{Action in online services.}}\label{subsub:online} 
\noindent In applets where the action (outcome) of an applet is transmitted to an online service (e.g., an email, a tweet, a Google calendar event), participants rated their concerns with using the applets if these services were shared with others or allowed access to other people (e.g., a public Facebook page). In fact, this was one of the few contextual factors for which participants were concerned (\textit{Extremely} ($37\%$),  Moderately ($22\%$)) in the majority of applets.

\subsubsection{Time of day.}
\label{subsubsec:time of day}
Testing the differences between two regression models (null model and Time-model) shows that participants' concerns varied significantly for \emph{time} ($\chi^2(df=2)=21.41, p<0.001$). The regression result (Time-model in Table~\ref{tab:glmm}) shows that participants were more concerned about the night time compared to morning. Pairwise comparisons highlight that participants were more concerned about night time when compared to afternoons ($p <0.05$, See Table~\ref{tab:timeglht} in the Appendix).

\subsubsection{Access.}
Security and privacy implications can arise because of \textit{who can use} the device (access the trigger) or \textit{who is around} to observe (access) the action. 
 
\noindent\textit{\textbf{Who can use.}}\label{subsubsec:whouse}
The regression analysis shows that ``who can use" was a significant factor in participants' concerns ($\chi^2(df=3)=338.27, p < 0.001$). Table~\ref{tab:glmm} (WhoCanUse-model) shows that participants were less concerned if their spouse would use these applets compared to other users. Overall, participants were less concerned about their family members using the applet---spouse ($Mean=2.29, SD=1.47$) and kids ($Mean=2.47, SD=1.47 $)---as compared to those who do not reside in the home---visitors ($Mean=2.7, SD=1.53$) and outsiders ($Mean=3.07,SD=1.58$). To find if these differences were significant, we ran pairwise comparisons and found that significant differences exist between all pairs of user types ($p <0.05$ for all pairs, see Table~\ref{tab:Whoglht} in the Appendix).

\noindent\textit{\textbf{Who is around.}}  The regression analysis shows that ``who is around" to observe the action event was a significant factor ($\chi^2(df=3)=71.727, p <0.001 $). Table~\ref{tab:glmm} (WhoIsAround model) shows that participants were more concerned about visitors and outsider as compared to their spouse. Pairwise comparisons found differences to be significant for outsiders and everyone else, as well as between visitors and spouse ($p<0.05$, Table~\ref{tab:Whoaglht} in the Appendix). This suggests participants did not differentiate much between family members when considering this contextual factor.

\begin{table}[t]
  \begin{center}
    \caption{Pairwise comparisons of different values for  Trigger location related to regression model TriggerLocation-model 2} 

    \label{tab:TLglht}
 \small
  
    \begin{tabular}{p{3cm}| p{1cm} p{1.3cm} p{1cm}  p{1.3cm}}
    
      \hline
     
     &Estimate & Std. Err. &z value &Pr($>|z|$)    
      \\
      \hline
      semi-private - private  &  -0.03821  &  0.03005 & -1.271 &  0.4097 \\
  public  - private  & 0.05193&    0.03583  & 1.449 & 0.3141\\
  public - semi-private &  0.09014   & 0.03607 &  2.499&   \textbf{0.0331}  \\ 
      
\hline

    \end{tabular}
   
  \end{center}
      \vspace{-1\baselineskip}
\end{table}

\subsection{Reasons to be concerned with using the applets in different usage context}

The quantitative analyses showed that participants' concerns increased almost universally when they considered the contextual factors. Here we explore the reasons for participants' concerns through qualitative analyses of the open-ended responses. Table \ref{tab:reasons} shows the top reasons across the six contextual factors.

\noindent \subsubsection{Leakage of sensitive data.}
About half of all open responses included concerns about users' sensitive data being collected and used in different contexts (See Table~\ref{tab:reasons}, last column). Participants were particularly concerned with applets connecting to \textit{online services} that could record their private information. For example, participant {P265} was concerned about a smart thermostat: ``\textit{... [others] will be able to tell when the thermostat is set to away mode, meaning no one is home}". A majority of concerns (97.26\% of valid responses for Action in Online Services, Table~\ref{tab:reasons} ) 
were about online logs of such sensitive information about user activities and schedules. 
ode, meaning no one is home.
Participants were also concerned about the sensitive nature of the data that could be recorded in private locations ($44.29\%$ of valid responses for \emph{Trigger Location}). As P319 reported: \textit{``certain rooms are sensitive and supposed to be private. I would not want it to collect info about people in the bathroom or bedroom, for instance."} They were also concerned about the visibility of the actions (24.82\% valid responses for \emph{Action Location}), especially when in semi-private or public places. For example, consider (Applet\#7) that enables printing a personal shopping list through Alexa. If the printer is \textit{located} in a semi-private space (e.g., living room), there is a chance that visitors (e.g., maid or babysitter) can also see the shopping list (and any sensitive items on the list). Participants reported concerns with who could view the actions (55.9\% responses for \textit{Who is Around}), as visitors or outsiders could then infer participants' schedules by noticing the actions. For example, P46 was concerned about (Applet\#4) that turns on the coffee maker when the user wakes up: \textit{``Because then ill intentioned people will learn my schedule and learn when I'm sleeping or awake."} Participants thought that the \textit{Time} of the day could impact the severity of the leakage  (60.75\%). For instance, P14 when talking about (Applet\#1) that posts an update status of users' location to Twitter said: \textit{``...I don't want where I am to be tweeted regardless of the time of day. But night time presents its own unique risks..."}

In addition to the data leakage because of the IFTTT applets, participants were also concerned about voice assistants recording private conversations without consent. As participant P332 reported: \textit{``I don't want my children or unsuspecting guests recorded without their permission."} Finally, some participants were concerned with leakage of details about how their home automation worked; participant P25 mentioned: \textit{ ``If someone overhears me turn off the camera, then it could be used without me knowing."}.

\noindent \subsubsection{Unauthorized/unintended access}

The second largest category of participants' concerns with applets involved unauthorized or unintended access to the applets and the resulting implications (Table \ref{tab:reasons}, last column). Like with leakage of sensitive data, \emph{location} of the IoT device raised concerns about unauthorized or unintended access. For example,  with (Applet\#5) that allows user to turn off the camera with a voice assistant like Alexa, {P34} reported their concern about Alexa being in semi-private locations such as the living room: \textit{``Because if its close to a visitor or outsider they can disable the cameras by saying the words to turn the cameras off."} Similarly, participants reported their concern with the risk of an applet controlled IoT device being in more sensitive areas. For instance, P289 responded: \textit{``... I don't want the bathroom light to be controllable from outside the bathroom."} For the applet that connected cameras, participants were concerned when the camera was located in private areas (bedrooms, bathrooms) as well as at the front door. As participant P2 reported: \textit{``I think the front[door] camera is likely the most important, so I don't want that turned off without my knowledge."} 

While unauthorized/unintended accesses were of significant concern when considering location of devices ($16.37\%$ of valid responses for \emph{Trigger Location}, $30.24\%$ of valid responses for \emph{Action Location}; See Table \ref{tab:reasons}), they were the single largest concern ($52.78\%$) when considering who is able to use the applets. In particular, participants highlighted their concern with having kids access to the applets, especially when those applets involved a voice assistant (e.g., Alexa) controlling IoT devices or online services. Participants reported that this was a concern because a child could play with Alexa and end up changing the settings for applets that control the lights (Applet \#36), thermostats (Applet \#34), or security cameras (Applet \#5). Such incorrect changes could in turn lead to incorrect configurations that could increase costs or put the house at risk. As participant P258 mentioned \textit{''kids can't touch it as they don't know how to properly use it. It needs to be set at a certain temp as I am on a savings plan. Would cost me money if others touched it."} Other examples included situations where a child could inadvertently say things that could be posted as tweets (Applet \#33), included in shopping lists (Applet \#7), or added as events to Google Calendar (Applet \#48). For example, participant P359  wrote about how Alexa could be misused:\textit{``My kids could ask Alexa something [unsuitable] or order things without permission."} Another participant P267 was wary of an applet that allows Alexa to post tweets ``\textit{Kids can say some stupid stuff sometimes, ....''} 

Apart from children, participants were also concerned about visitors and outsiders gaining access to their applets, particularly those related to security features such as security camera in the case of visitors. For outsiders, their concerns encompassed almost all applets. Participant P14 quoted: \textit{``I am not concerned with who can turn the switch on, unless they are an outsider. Outsider implies [someone] I have not invited into my home."}

\begin{table}
\small
\begin{threeparttable}
  \caption {Top coded reasons for being concerned to use applet in each usage context}
  \label{tab:reasons}
  \begin{tabular}{p{3.7cm}| p{1cm}p{1cm}p{1cm}|p{.7cm}|p{.9cm}p{.9cm}|p{1cm}}
    \toprule
   & \multicolumn{3}{c|}{Location}&Time& \multicolumn{2}{c|}{Access}&\\
   \cline{2-7}
    Reasons & Trigger location & Action location & Online Services\tnote{1}{} & Time& Who can use& Who is around&\%of all\newline responses \\
    \midrule
    
     Leakage of sensitive data & 44.29\% &24.82\%&97.26\%&60.75\%&8.97\%&55.90\%&49.50\%\\
    Unauthorized/unintended access & 16.37\%&30.34\%&-&6.32\%&52.78\%&18.50\%& 22.89\%\\
     Privacy/Security misconception & 17.25\%&3.44\%&1.37\%&3.79\%&34.13\%&4.72\%&14.13\%\\
       Inconvenience & 11.06\%&30.34\%&-&12.44\%&-&6.69\%&5.77\%\\
    Safety & 2.65\%\% &4.82\%&-&5.90\%&-&8.66\%&2.50\%\\

  \bottomrule

\end{tabular}
 \begin{tablenotes}
      \item[1]\small{
      Action in online services}
     
   \end{tablenotes}
\end{threeparttable}
\end{table}
 \noindent \subsubsection{Privacy/Security misconception.}
  Several participants' responses ($14.13\%$ of all responses) also highlighted misconceptions about privacy and security risks. Most of these are related to applets that connect IoT devices (e.g., security cameras and smart locks), or location to online services. In particular, participants could not distinguish between unauthorized/unintended modification (write access) to their online accounts and leakage of information from those accounts (read access). For instance, in an applet that sends an email to the user when the security system is turned off (Applet\#15), 
  participant {P145}  reported: \textit{``
  Because I worry about people getting access to my email and turning off the system.}
\noindent \subsubsection{Inconvenience.}
About 6\% (Table~\ref{tab:reasons}, last column) of responses included concerns about the inconvenience of using the applets in different contexts. A majority of these were about the \textit{Action Location} (30.34\%), especially for devices that notified users through blinking or turning on lights as they could disturb others during the night (Time, 12.44\%, Who is Around: 6.69\%). For instance, participant P181 said: \textit{``If it blinks the lights in my bedroom it could wake me up"}.
In 11.06\% cases, participants were concerned about the accessibility of the (Trigger) location, especially the placement of voice assistants in the house, as  
participant P163 reported: \textit{``if my voice does not reach the Alexa it does not work."}

\noindent \subsubsection{Safety.} A small portion of concerns (2.5\%, Table~\ref{tab:reasons}) were about the safety of using smart appliances, such as smart coffee makers, ovens, and vacuums. For example, participant P41 was worried if (Applet\#4) would start the coffee maker when they wake up but still in bed and if their child was unsupervised in the kitchen: \textit{``My child, because I do not want the coffee to start brewing and she touches it."} On the other hand, some of the concerns were largely about unsupervised operation of smart appliances regardless of them being connected via IFTTT applets. For example, participant P133 was concerned about (Applet\#17) that enables user to turn off the oven through Alexa: \textit{``I would be concerned about how busy I am and if I would be able to pay attention to the oven if its turned on after I activated it. I might be busy with errands/chores at home."}

\section{Discussion} 

\boldification {sharing on online services matters but reviewing responses shows that participants need more information about context regarding with whom data is shared. }

Our findings indicate that the descriptions of trigger-action applets were insufficient for participants to assess the risks of using IFTTT applets. Participants on average had low concerns with using these applets except for when the location data were reported to online services. But, when prompted to think about specific contexts they almost universally increased their concerns. In these cases, their main concerns (72.39\%) were largely about leaking sensitive data and unauthorized or unintended access. Our study is the first validation survey that shows the value of context when determining security and privacy risks in IFTTT applets for home automation and opens the field for two future research opportunities.

\subsection{Help end users assess security and privacy risks}

Our study found two aspects where participants struggled with assessing the security and privacy risks associated with IFTTT applets.
First, in our survey, even with explicit trigger-action descriptions, participants reported low concerns with using applets with potential security and privacy risks. Of course, one can argue that this low concern could be an artifact of the study setup itself, where participants ``simply" answered the survey question, whereas in real life they might have given a deeper consideration to the descriptions. However, past research~\cite{zheng2018user} has also shown that end users have difficulty in understanding security and privacy implications of IoT devices in some cases.  
When prompted to think specifically about the contextual factors, participants universally raised their concern scores. This shows that when nudged participants could identify the risks associated. However, participants spent an average of 3-4 minutes per applet answering the baseline question and thinking about the contextual factors (recall, each survey was about 30 min long and included questions about 7 applets). Although 3-4 minutes is not much, end users may not spend that long to preview contextual information when they install applets. Studies have shown that users, when installing applications, just want to get the application running and rarely change default settings or read privacy disclosures~\cite{Bohme2010, Schaub2015}.

Second, in many cases participants had difficulty understanding the security \& privacy risk implications. For example, while participants readily understood the risks of leaking their location data to online services (recall the applets to the left of Figure~\ref{"mean"}), they were far less concerned about applets that indirectly leaked user's presence or location by sharing house's thermostat configurations (applets in right end of Figure~\ref{"mean"}). Moreover, the analysis of the open-ended responses revealed there were security \& privacy misconceptions (14.3\%) across all the contextual factors, which were primarily related to applets connecting to online services. This is likely because of mismatches in participants' mental models of what and where the applet records and the reality of how the applet operates. Past work~\cite{kang2014privacy,KaazHSSB17} has shown that end users, especially those without technical background, have simplistic mental models about information flows and who had access to their personal data or communication.

These findings present future research opportunities on how we can facilitate end-users' security \& privacy risk assessments to help them make informed choices. It is an open question of what kinds of intuitive designs can applet designers (or platforms like IFTTT) create that allow end users to easily and efficiently reflect on the different contextual factors that are applicable to a particular applet.  Moreover, the contextual factors are likely to be more useful if they are personalized to the users' specific needs and lifestyle.

One option is giving a ``nudge" to the users about the potential security and privacy risks associated with the applet, similar to the nudges proposed by~\cite{masaki2020exploring} for sharing online information in social network services.
Future research should investigate mechanisms to automatically create such security \& privacy nudges based on the contextual factors applicable to a given applet and tailored to a users' profile or past data usage. Another lighter-weight option could be a method to identify a set of standardized contextual factors for each class of applets and some metrics for portraying confidentiality (leakage of sensitive data) and integrity (unauthorized use) violations, akin to the concept of standardized labels proposed by Kelley \etal~\cite{Kelley2010} for comparing privacy policies of websites. 
 
\subsection{Incorporate nuanced context in the design of automated approaches}

Our work opens the door for further research on the interactions between the contextual factors as well. In this work, we have only looked at a handful of contextual factors---those that were applicable to the 49 applets we studied. There may be other contextual factors that are relevant for other types of applets. Further, the different contextual factors interact. For example, a camera that can be switched off is much more problematic, if it is a security camera, if it is switched off at night time, and there are (young) kids in the house. Such interaction of different contextual factors and the implication of these interactions on privacy concerns needs further study.

\section{Conclusion}
In this paper we reported on a large-scale Mechanical Turk survey (n=386) on users'
concerns related to \textit{trigger-action} applets for smart homes that are available through the IFTTT platform. 

Our results enhance the findings of previous works by showing how different contextual factors affect end-users' concerns. Our analysis shows that even when descriptions explicitly mention the trigger-action components in applets, participants fail to perform adequate risk assessments, expressing low concerns with using applets with underlying security and privacy threats.

Our study shows that contextual factors are nuanced and their interplay affects users' concerns. We have only scratched the surface---evaluating 6 contextual factors--- and shown that research opportunities exist to better understand \ci the role that context plays in end-users' concerns and \cii how to devise mechanisms to help end users evaluate the context of use better. The extreme 
popularity and ubiquity of trigger-action applets necessitates further investigations into how we can help end users better perform risk assessments of trigger-action applets.

\bibliographystyle{plain}
\bibliography{ms.bib}
\newpage

\appendix
\section{Applet Descriptions}

\begin{longtabu} to \textwidth {|p{0.8cm}|p{13cm}|}
 \caption{Applet descriptions\label{long}}\\

 \hline
 
 App\# & \centering Applet description\\
 \hline
 \endfirsthead

 \hline
 
 App\# & \centering  Applet description\\
 \hline
 \endhead

   App1 & This applet connects your location and Twitter. This specific applet will post a tweet to twitter when you enter a location. \\  \hline
App2 & This applet connects your WeMo Insight Switch and email. This specific applet sends an email every time your switch is turned on.\\  \hline
App3  & This applet connects your EverNote and location. This applet specifically will add a new note when you enter an area.\\ \hline

App4 & This applet connects your Fitbit and WeMo coffee maker. This specific applet turns your coffee maker on when your fitbit logs that you have woken up for the day.\\ \hline
App5 & This applet connects your Alexa and Nest security camera. This specific applet will turn your camera off when you say "Alexa, turn off camera".\\ \hline

App6 & This applet connects your Nest security camera and Google spreadsheet. This specific applet will add a new row to your spreadsheet when your camera detects motion. \\
\hline
App7 & This applet connects your printer and your Alexa. This specific applet will print your grocery list when you say "Alexa what's on my shopping list". \\ \hline

App8 & This applet connects your Samsung Robot vacuum and time. This specific applet has your vacuum start cleaning everyday at a specific time. \\  \hline

App9 & This applet connects your Alexa and your appliance. This specific applet will turn your TV on when you ask Alexa to "turn on your TV". \\ \hline

App10 & This applet connects your location and EverNote. This applet specifically will add a new note when you exit an area.\\ \hline

App11 & This applet connects your WeMo plug and google calendar. This specific applet adds an event to your calendar every time your switch is turned on.\\  \hline

App12 & This applet connects your email and WeMo light switch. This specific applet sends you an email daily giving you the details of how much it  costs to operate your light switch. \\  \hline

App13 & This applet connects your location and Facebook. This specific applet will post a Facebook status when you enter an area. \\ \hline

App14 & This applet connects your Nest thermostat and time. This specific applet turns your thermostat on once a day at a specific time for the length of time you decide. \\ \hline

App15 & This applet connects your Arlo security system and email. This specific applet will send you an email when your security system is turned off.\\ \hline

App16 & This applet connects your Google spreadsheet and location. This specific applet will add a row to your spreadsheet when you exit an area. \\ \hline

App17 & This applet connects your appliance and Alexa. This specific applet will turn off your oven when you ask Alexa to "turn off your oven".\\ \hline

App18 & This applet connects your Ring doorbell and EverNote. This specific applet adds a note when your doorbell detects motion as a visitor log. \\ \hline

App19 & This applet connects your location and email. This specific applet will email you when you leave an area.\\  \hline

App20 & This applet connects your Phillips Hue lights and Android SMS. This specific applet will blink your lights when you receive an SMS. \\ \hline

App21 & This applet connects your Alexa and EverNote.This specific applet will add items to your list when you say "Alexa add milk to my grocery list". \\ \hline

App22 & This applet connects your location and Phillips Hue lights. This specific applet will turn off all the lights when you leave a specific area. \\
\hline

App23 & This applet connects your Samsung washer and EverNote app. This specific applet adds a note when you start a new washer cycle.\\ \hline

App24 & This applet connects your location and Google spreadsheet. This specific applet will add a row to your spreadsheet every time you enter an area.\\ \hline

App25 & This applet connects your Alexa and Phillips Hue lights. This specific applet will turn your lights turn off when you ask Alexa to "turn lights off".\\ \hline

App26 & This applet connects your Samsung washer and Phillips Hue lights. This specific applet will blink your lights when your washer has started a load.\\ \hline

App27 & This applet connects your Alexa and Nest thermostat. This specific applet will adjust your temperature when you say "Alexa, set thermostat to 62 degrees". \\ \hline

App28 & This applet connects your Ring doorbell and Google calendar. This specific applet will add an event to your calendar when your doorbell detects motion.\\ \hline

App29 & This applet connects your  Samsung washer and google calendar. This specific applet adds an event to your calendar when the washer has finished a wash cycle. \\ \hline

App30 & This applet connects your Nest thermostat and email. This specific applet will send you an email when your thermostat set away. \\ \hline

App31 & This applet connects your email and Kevo lock. This specific applet will send you an email when your lock is locked.\\ \hline

App32 & This applet connects your Phillips Hue lights and time. This specific applet will turn your lights off at a specific time everyday.\\ \hline

App33 & This applet connects your  Alexa and Twitter. This specific applet will post a tweet to your twitter when you say "Alexa, post a tweet".\\ \hline

App34 & This applet connects your  Alexa and Nest thermostat. This specific applet will turn your fan on for 15 minutes when you say "Alexa, turn fan on".\\  \hline

App35 & This applet connects your Nest security camera and email. This specific applet sends you an email when your camera detects motion.\\ \hline

App36 & This applet connects your Alexa and Phillips Hue lights. This specific applet will turn your lights on when you ask Alexa to "turn on lights".\\ \hline

App37 & This applet connects your email and SmartThings hub. This specific applet will send you an email when it detects motion. \\ \hline

App38 & This applet connects your Wemo plug and email. This specific applet sends an email when your plug is turned on.\\ \hline

App39 & This applet connects your email and Kevolock. This specific applet will send you an email when your lock is unlocked. \\ \hline

App40 & This applet connects your Alexa and Spotify. This specific applet will play your Spotify when you say "Alexa, play my Spotify top hit playlist".\\ \hline

App41 & This applet connects your location and email. This specific applet will send you an email when you enter an area. \\ \hline

App42 & This applet connects your Ring doorbell and Phillips Hue lights. This specific applet will blink your lights when your doorbell detects motion. \\ \hline

App43 & This applet connects your Phillip Hue lights and Alexa. This specific applet will blink Philips Hue lights when you say "blink the lights".\\ \hline

App44 & This applet connects your Google spreadsheet and SmartThings hub. This specific applet will add a row to your spreadsheet when your hub detects your motion. \\ \hline

App45 & This applet connects your Samsung washer and email. This specific applets sends you an email when your washer has completed a cycle.\\ \hline

App46 & This applet connects your email and Nest thermostat. This specific applet will send you an email when your thermostat set home. \\ \hline

App47 & This applet connects your Arlo security system and email. This specific applet will send you an email when your security system has a problem. \\
\hline

App48 & This applet connects your  Alexa and Google calendar. This specific applet will add the next game to your calendar when you say "Alexa when do the Golden state warriors play next".  \\ \hline

App49 & This applet connects your location service on your smart phone and Arlo security system. This specific applet will turn your security system on when you leave your home.
\\ \hline
 \end{longtabu}

\section{Sample Survey questions}
 
 \subsection{ Demographics questions}
 \begin{enumerate}

\item What is your gender?
\begin{itemize}
    \item Male
    \item Female
    \item Trans 
    \item Non-binary 
    \item Prefer not to say
\end{itemize}
\item What is your age?
 
 \begin{itemize}
     \item  18-24
\item 25-34
\item 35-44
\item 45-54
\item 55-64
\item 65+
\end{itemize}
\item What is the highest level of education you have completed?
\begin{itemize}
    \item Less than high school
    \item High school graduate
\item Some college
\item 2 year degree
\item 4 year degree
\item Professional degree
\item Doctorate,
\item Prefer not to answer
\end{itemize}
\item How many IFTTT applets have you used?
\begin{itemize}
\item 0
\item 1-2
\item 3-4
\item More than 5
\end{itemize}
\item How many smart home devices do you own?
\begin{itemize}

\item 0
\item 1-2
\item 3-4
\item More than 5
    
\end{itemize}
 \end{enumerate}

\subsection{ Applet questions}
 
 \textbf{Applet description:}
Suppose there is an applet that connects your \textbf{Alexa} and your \textbf{Nest security camera}. This specific applet will turn your camera off when you say "Alexa, turn off camera". 
 \begin{enumerate}
     \item 
Would you be concerned about using this applet? (Answered on a five point Likert scale from “Not at all concerned” to “Extremely concerned”)

\item To what extent do the following factors affect your concern in using this applet?  (Answered on a five point Likert scale from “Not at all concerned” to “Extremely concerned”)
\begin{itemize}
    \item The time when you ask Alexa\\
    (Morning, Afternoon, Night)
    \item The location of Alexa
    
    (Living room, Kitchen , Bedroom, Bathroom)
    \item Who can talk to Alexa 
    
    (Spouse, Kids, Visitor, Outsider)
    
    \item The location of camera		
    (Living room, Kitchen , Bedroom, Bathroom, Frontdoor)
    \item Who is around
    
    (Spouse, Kids, Visitor, Outsider)
    
\end{itemize}
\item Why are you concerned with \textbf{what time} you talk to Alexa?
\item Why are you concerned with the \textbf{location of your Alexa}?
\item Why are you concerned with \textbf{who can talk} to your Alexa? 
\item Why are you concerned with \textbf{who is around}?
\item Why are you concerned with the \textbf{location of your camera}?

\end{enumerate}
\section{Pairwise comparisons}

\newpage
\begin{table}[t]
  \begin{center}
    \caption{Pairwise comparisons of different values for  Action location (See Section \ref{subsubsec:loc})}

    \label{tab:alocglht}
 \small
  
    \begin{tabular}{p{3cm}| p{1cm} p{1.3cm} p{1cm}  p{1.3cm}}
    
      \hline
     
     &Estimate & Std. Err. &z value &Pr($>|z|$)    
      \\
      \hline
      private - semi-private &  0.1230& 0.0293&   4.201 & \textbf{0.0001} \\
 private - public    & $-0.0219$ & 0.0345 &  $-0.636$ &  0.8004 \\
 semi-private - public &  $-0.1450$ & 0.0353&  $-4.112$ &  \textbf{0.0001} \\ 
      
\hline

    \end{tabular}
   
  \end{center}
      \vspace{-1\baselineskip}
\end{table}
\begin{table}[h]
  \begin{center}
    \caption{Pairwise comparisons of different values for  Time (See Section \ref{subsubsec:time of day})
    }
    \label{tab:timeglht}
 \small
  
    \begin{tabular}{p{2.7cm}| p{.8cm} p{1.1cm} p{1cm}  p{1.2cm}}
    
      \hline
     
     &Estimate & Std. Err. &z value &Pr(>|z|)    
      \\
      \hline
      Afternoon - Morning & 0.007 &  0.017 &  0.452 &0.893    
      \\
Night - Morning &   0.072 &  0.017 &  4.189  & \textbf{<0.001} 
      \\
Night - Afternoon &  0.064  & 0.017 &  3.738 & \textbf{0.0005} 
           \\
      
\hline

    \end{tabular}
   
  \end{center}
      \vspace{-1\baselineskip}
\end{table}

\begin{table}[h]
  \begin{center}
    \caption{Pair wise comparisons of different values for  Who can use
    (See Section \ref{subsubsec:whouse})}
    \label{tab:Whoglht}
 \small
  
    \begin{tabular}{p{2.5cm}| p{.8cm} p{1.1cm} p{1cm}  p{1.2cm}}
    
      \hline
     
     &Estimate & Std. Err. &z value &Pr(>|z|)    
  \\
      \hline
Kids - Spouses &0.049 &    0.018&   2.593&  \textbf{ 0.046}   
      \\
Visitors - Spouses &  0.188 &   0.018&  10.278 &  \textbf{<0.001} 
      \\
Outsiders - Spouses &  0.294&    0.017&  16.428&   \textbf{<0.001 }
           \\
           Visitors - Kids  &     0.139  &  0.018&   7.699   & \textbf{<0.001}
\\
Outsiders - Kids      & 0.245&    0.017 & 13.87&   \textbf{<0.001} 
\\
Outsiders - Visitors&   0.105  &  0.017  & 6.222 &  \textbf{<0.001}

    \end{tabular}
   
  \end{center}
      \vspace{-1\baselineskip}
\end{table}

\begin{table}[h]
  \begin{center}
    \caption{Pairwise comparisons of different values for  Who is around (See Section \ref{subsubsec:whouse})
    }
    \label{tab:Whoaglht}
 \small
  
    \begin{tabular}{p{2.5cm}| p{.8cm} p{1cm} p{1cm}  p{1.2cm}}
      \hline
     & Estimate & Std. Error &z value & Pr(>|z|)    
      \\
      \hline
Kids - Spouses &        0.037  & 0.027&   1.367 &  0.519   
      \\
Visitors - Spouses& 0.102 &   0.026&   3.815&   \textbf{<0.001} 
      \\
Outsiders - Spouses &  0.206&    0.026  & 7.883  & \textbf{<0.001} 
\\
Visitors - Kids &   0.065 &   0.026 &  2.450 &  0.068   
\\
Outsiders - Kids     &  0.169  &  0.025 &  6.527  & \textbf{<0.001 }
\\Outsiders - Visitors &  0.104&    0.025 &  4.087&   \textbf{<0.001} 
    \end{tabular}
   
  \end{center}
      \vspace{-1\baselineskip}
\end{table}

\end{document}